\documentclass[
 preprint,
 superscriptaddress,
 amsmath,amssymb,
 floatfix
]{revtex4-1}

\usepackage[english]{babel}
\usepackage{graphicx}
\usepackage{upgreek}
\usepackage{bm}
\usepackage{textcomp}

\begin{document}

\title{Strong and Weak Three-Dimensional Topological Insulators Probed by Surface Science Methods}

\author{ Markus Morgenstern}
\email{mmorgens@physik.rwth-aachen.de}
\author{  Christian Pauly}
\author{  Jens Kellner}
 \author{  Marcus Liebmann}
 \author{  Marco Pratzer}
\affiliation{II. Institute of Physics B and JARA-FIT, RWTH Aachen University, 52074 Aachen, Germany}
\author{   Gustav Bihlmayer}
\affiliation{Peter Gr\"{u}nberg Institut (PGI-1) and Institute for Advanced Simulation (IAS-1), Forschungszentrum J\"{u}lich GmbH and JARA, 52425 J\"{u}lich, Germany}
\author{  Markus Eschbach}
\author{  Lukacz Plucinski}
\affiliation{Peter Gr{\"u}nberg Institut (PGI-6), Forschungszentrum J{\"u}lich GmbH and JARA, 52428 J{\"u}lich, Germany}
\author{  Sebastian Otto}
 \affiliation{Lehrstuhl f{\"u}r Festk{\"o}rperphysik, Universit{\"a}t Erlangen-N{\"u}rnberg, 91058 Erlangen, Germany}
 \author{  Bertold Rasche}
\author{  Michael Ruck}
\affiliation{Faculty of Chemistry and Food Chemistry, TU Dresden, 01062 Dresden, Germany}
\author{  Manuel Richter}
\affiliation{Leibniz Institute for Solid State and Materials Research, IFW Dresden, P.O. box 270116, 01171 Dresden, Germany and Dresden Center for Computational Science}
\author{  Sven Just}
\author{  Felix L\"upke}
\author{  Bert Voigtl\"ander}
\affiliation{Peter Gr{\"u}nberg Institut (PGI-3), Forschungszentrum J{\"u}lich GmbH and JARA, 52428 J{\"u}lich, Germany}

\date{\today}


\maketitle

\textbf{We review the contributions of surface science methods to discover and improve 3D topological insulator materials,
while illustrating with examples from our own work. In particular, we demonstrate that spin-polarized angular-resolved photoelectron spectroscopy
is instrumental to evidence the spin-helical surface Dirac cone, to tune its Dirac point energy towards the Fermi level, and to
discover novel types of topological insulators such as dual ones or switchable ones in phase change materials.
Moreover, we introduce procedures to spatially map potential fluctuations by scanning tunneling spectroscopy and to identify topological edge states
in weak topological insulators.
}
\section{Introduction}
Topology became a classification scheme for solid state electronic properties in the 1980s while describing the robustness of the quantum Hall effect \cite{Thouless1982,Niu1985}. This achievement has been honored most notably by the Noble prize 2016 for physics \cite{Cho2016,denNijs2019}.
The well-deserved appreciation was largely triggered by the experimental discovery of 2D topological insulators (2DTIs)  in 2007 \cite{Konig2007}. This discovery initiated a major effort in experimental and theoretical solid state physics leading to
a multitude of other types of topologies in crystalline solids, mostly appearing without magnetic fields \cite{Hasan2010,Qi2011,Ando2013}. The overwhelming success has also led to activities in other fields of physics enabling, e.g., the guiding of light or sound along arbitrarily shaped edges \cite{Bahari2017,Cha2018,Yang2019,Ningyuan2015}.
The attractive robustness of the topological properties, tied to the integer character of the topological indices, implied a multitude of proposals also
for electronic applications \cite{Zhang2010,He2013,Han2017}. This currently culminates in the actively pursued dream to realize topological quantum computation via parafermions \cite{Alicea2012,Sarma2015,Alicea2016,Lahtinen2017}. The central advantage of this approach is
the robustness of corresponding quantum operations against local perturbations as long as the quasiparticles remain in their topologically protected subspace.

From the point of view of materials science, the intriguing observation that a lot of well-known materials are
three-dimensional strong topological insulators added a crucial view on electronic band structure properties \cite{Hasan2010,Qi2011,Ando2013}.
It turned out that a large amount of bulk insulators necessarily provide spin helical conductive surface states \cite{Zhang2019,Vergniory2019} via the symmetry
of their bulk band structure described by a topological index \cite{Moore2007,Fu2007b}.
The presence of such surface states is totally independent on details of the confining surfaces and, moreover,
these surface states are protected against backscattering by their spin helicity \cite{Fu2007,Hasan2010}. Hence, such materials can be thought of as a third conductivity class
besides conductors and insulators, being insulating in the interior of the system but conducting on its surfaces. Favorably, a simple classification scheme exists in case of inversion symmetry of the crystal \cite{Fu2007}. It simply multiplies the parities (point inversion symmetries) of occupied single-electron states at the time-reversal invariant momenta (TRIMs) of the Brillouin zone in order to deduce the topological index. This provides an easy tool to exploit the much more complex theoretical background, that relies on extracting
topological indices from general symmetries of the describing Hamiltonian \cite{Moore2007,Fu2007b,Kane2005}. High-throughput density functional theory (DFT) calculations can be used to automatically extract candidate topological insulators from the extensive data base existing for crystalline materials \cite{Zhang2019,Vergniory2019}. This often leads to materials with large band gap such that the topological transport properties can be observed at room temperature.

However, subsequently, the candidate materials still have to be verified and characterized by experimental methods. This is due to the inherent minor difficulties of DFT calculations such as
the missing precise description of electronic correlations and of van-der-Waals interactions \cite{Mardirossian2017,Stoehr2019} as well as the typically too small band gap. Since surface states are the decisive fingerprint of 3D topological insulators,
well-established surface science methods became the tool of choice for the task of confirmation. In particular, angular resolved photoelectron spectroscopy (ARPES) directly maps the spin helical surface states in $\bm{k}$ space \cite{Hsieh2008,Hsieh2009}, that typically exhibit a Dirac-type linear dispersion around one of its TRIMs in the Brillouin zone \cite{Zhang2009}.
The spin-polarized version of ARPES (SARPES) moreover can characterize the spin-helical Dirac character of the topological surface states \cite{Hsieh2009}. Both can be compared directly with DFT based calculations enabling an immediate verification of the topology \cite{Hasan2010,Qi2011,Ando2013}. Moreover, the doping level and, thus, the position of the intrinsic Fermi level $E_{\rm F}$ with respect to the Dirac point energy $E_{\rm D}$ can be checked via ARPES \cite{Hsieh2009b}. This is decisive for any type of applications in electronic devices requiring the Dirac cone to be present at $E_{\rm F}$.

For exploitations of topological insulators in electric transport experiments, it turned out that disorder is detrimental \cite{Culcer2012}. Firstly, point defects acting either as acceptors or donors can make the interior conductive by shifting $E_{\rm F}$ into bulk bands \cite{Yan2012,Cheng2010}. Hence, the bulk conductivity often overwhelms the conductivity of the topological surface states \cite{Devidas2014}. Secondly, surface doping can lead to a surface band bending that hosts additional 2D states of non-topological origin at $E_{\rm F}$, while the topological surface states (TSSs) are detuned from $E_{\rm F}$ \cite{Bianchi2010,King2011}. The latter is difficult to avoid, since any contamination on the surface, resulting, e.g., from device preparation, can imply a band bending that even appears after a few minutes of UV illumination in ultrahigh vacuum (UHV) \cite{Frantzeskakis2017}. Finally, even in case that $E_{\rm F}$ is favorably positioned within the bulk band gap, compensation doping can lead to such strong potential fluctuations that electron and hole puddles appear in the interior of the sample implying hopping transport that competes with the transport via the TSSs \cite{Skinner2012,Borgwardt2016}. Thus, experimental access to the potential disorder is crucial for improving the transport properties. The potential disorder can be mapped on small length scales by scanning tunneling spectroscopy (STS) \cite{Morgenstern2000,Morgenstern2002}. Therefore, one either employs the spatial variation of features in the local density of states (LDOS) related to $E_{\rm D}$ or the band edges \cite{Beidenkopf2011,Dai2016,Kellner2017} or, more precisely in energy, by spatially tracking Landau level energies in magnetic field $B$ \cite{Cheng2010,Hanaguri2010,Fu2014,Okada2012,Pauly2015}.
\begin{figure*}[tbh]%
\includegraphics*[width=\textwidth]{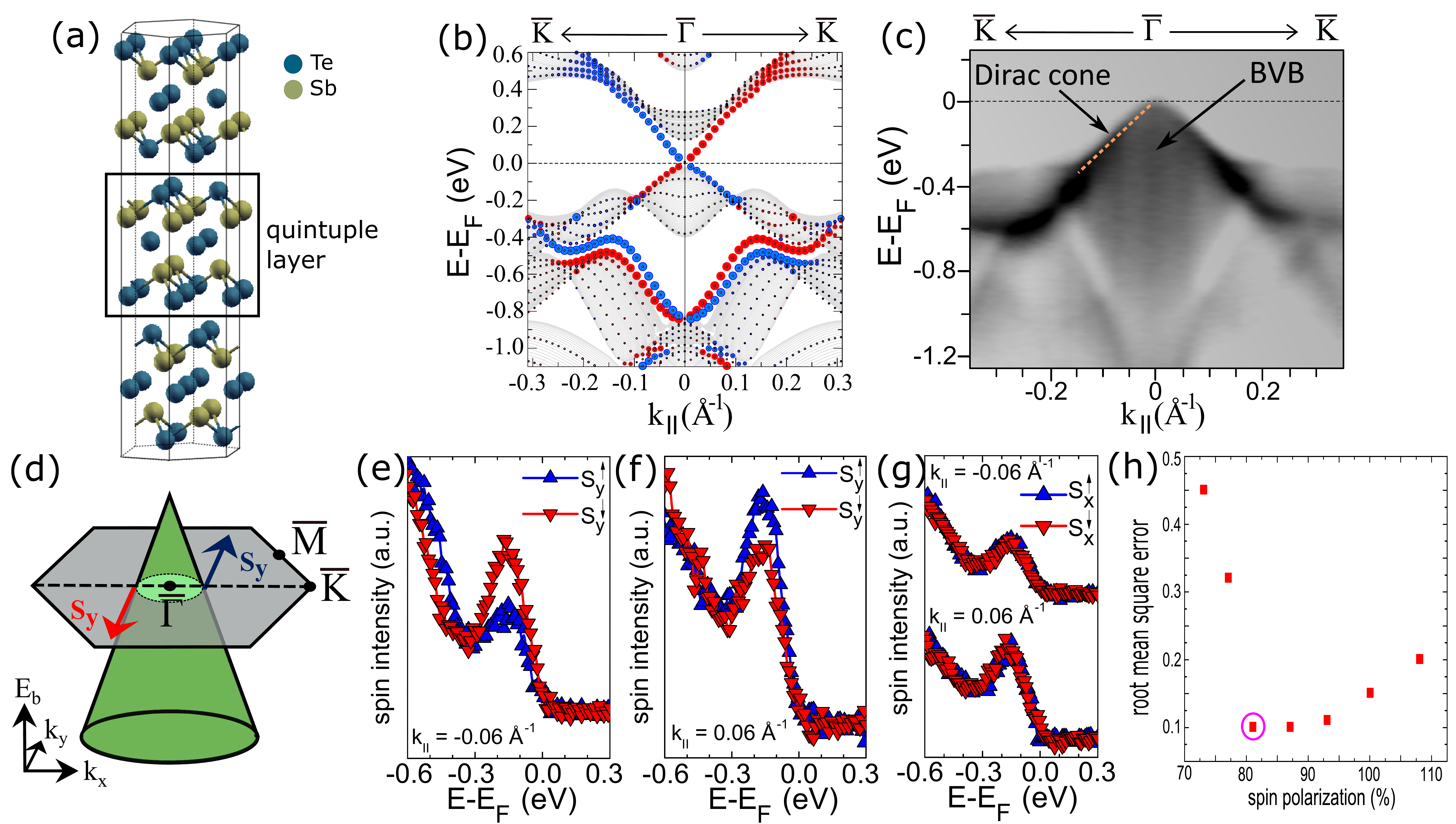}
\caption{{\it Identifying topological surface states.} (a) Structural model of Sb$_2$Te$_3$ with marked quintuple layer. (b) Band structure of Sb$_2$Te$_3$ in
 $\overline{\mathrm K}\overline{\mathrm \Gamma}\overline{\mathrm K}$ direction as calculated by DFT including spin-orbit coupling. States are shown as circles with colors (blue or red) that indicate different in-plane spin directions perpendicular to ${\bm k}_\parallel$ as resulting from a thin film calculation. The size of the colored circles marks the magnitude of the spin density of the state near the surface. Shaded
 areas are projected bulk bands originating from a bulk calculation. The strongly spin-polrized states have been checked to be surface states. (c) ARPES data of the lower Dirac cone of in-situ cleaved Sb$_2$Te$_3$(0001) recorded along  $\overline{\mathrm K}\overline{\mathrm \Gamma}\overline{\mathrm K}$ (dark: high intensity, bright: low intensity). States from the bulk
 valence band (BVB) are marked. Orange, dashed line is a guide to the eye along the Dirac cone revealing a Dirac velocity $v_{\rm F}=3.8\pm 0.2 \cdot 10^{5}$\,m/s, photon energy $h\nu=55$\,eV. (d) Sketch of the lower Dirac cone with spin directions $s_y$ marked as deduced from DFT (b) and in accordance with SARPES (e$-$g). (e), (f) Spin-resolved energy distribution curves (EDCs) for the spin component perpendicular to $\bm k_{\rm \|}$
at $k_{\rm \|}$-values as indicated. Different colors mark different spin directions as in (b), $h\nu = 54.5$ eV.
(g) Spin-resolved EDCs for the spin component parallel to $\bm k_{\rm \|}$,
$h\nu = 54.5$ eV. (h) Root-mean-square error for the deduced spin polarization of the topological surface state according to the SARPES data of (e), (f) after employing fits to
adequately subtract contributions form bulk bands, other surface bands and inelastic scattering (for details see \cite{Pauly2012}). Best fit value is encircled. $T=300$\,K \cite{Pauly2012}.}
\label{Fig1}
\end{figure*}
Additionally, STS can map 1D topological states that are difficult to probe via ARPES \cite{Takayama2015,Bianchi2015}, since these states are typically only sparsely dispersed on the surface. STS identifies the topological edge states straightforwardly as increased intensity of the LDOS at step edges \cite{Drozdov2014,Pauly2015b,Li2016,Wu2016,Sessi2016,Peng2017,Reis2017}. Its distinctive property of prohibited backscattering appears via the missing standing waves. Such standing waves are very pronounced for conventional 1D electronic states due to the strongly confined 1D geometry \cite{Meyer2003}. Hence, its absence is  a strong fingerprint of prohibited backscattering. 1D topological edge states have been found for 2D topological insulators \cite{Drozdov2014,Reis2017,Kim2016}, weak topological insulators \cite{Pauly2015b,Li2016,Wu2016} and at step edges of topological crystalline insulators, where they are caused by a symmetry breaking of the crystal at these edges \cite{Sessi2016}.

In this short review, we will exemplify the mentioned surface science based approaches to topology in crystals. These approaches are still central tools for the characterization of different topologies as well as for the finding of novel prospective materials within an established class of topology \cite{Bansil2016}. For the sake of simplicity, we restrict illustrations to our own work that cover many of the central developments yet.
We can not review the literature  extensively, already due to the bare amount, but concentrate on initial publications and central additional insights on methodology.

\section{Identifying topological surface states}

Soon after establishing 2DTIs experimentally \cite{Konig2007} based on theoretical predictions \cite{Kane2005}, an extension of the formalism to 3D was proposed \cite{Fu2007,Fu2007b}. It results
in two types of 3D topological insulators (3DTIs). One exhibits an odd number of spin-helical TSSs on each surface and is dubbed strong 3DTI, while the other one has an even number
of topological surface states on every surface except one and is dubbed weak 3DTI \cite{Fu2007,Fu2007b}. After identifying a first strong 3DTI in a BiSb alloy by ARPES \cite{Hsieh2008}, DFT calculations predicted stoichiometric materials to be strong 3DTIs, namely Bi$_2$Te$_3$, Sb$_2$Te$_3$, and Bi$_2$Se$_3$ \cite{Zhang2009}. These three materials share the same crystal structure of quintuple layers (QL) that are stacked on each other by van-der-Waals forces (Fig.~\ref{Fig1}(a)). Hence, these materials can be cleaved in-situ and can be exfoliated as thin films \cite{Lampert1981,Teweldebrhan2010}. Moreover, they have been
predicted to exhibit a single TSS on the cleavage plane with the Dirac point located in the center of the Brillouin zone at the so-called $\overline{\Gamma}$ point (Fig.~\ref{Fig1}(b)) \cite{Zhang2009}.

These properties enable a simple investigation by SARPES provided that the Dirac cone (TSS) is below $E_{\rm F}$. Indeed, the first ARPES measurements of a TSS on Bi$_2$Se$_3$(0001) have been published \cite{Xia2009} back-to-back with the DFT based predictions \cite{Zhang2009}. First SARPES measurements appeared only three month later \cite{Hsieh2009}. It turned out that the cleaved bulk samples of Bi$_2$Se$_3$ and Bi$_2$Te$_3$ are n-doped, being beneficial for the ARPES mapping of the TSS, but detrimental for electric transport. In contrast, Sb$_2$Te$_3$ is usually p-doped \cite{Hsieh2009c,Seibel2015} impeding ARPES mapping. Luckily, we obtained a twenty year old Sb$_2$Te$_3$ crystal that enabled mapping of the lower part of the Dirac cone via ARPES (Fig.~\ref{Fig1}(c)) \cite{Pauly2012}. This part of the Dirac cone encloses states of the bulk valence band in $\bm {k}$ space in quantitative accordance with DFT calculations (Fig.~\ref{Fig1}(b)). Since the doping is caused by point defects of the material \cite{Wang2010,Jiang2012}, we speculate that the particular defect distribution within this material is responsible for establishing the favorable $E_{\rm D}\simeq E_{\rm F}$. Similar results exhibiting Dirac cones within the band gap close to $E_{\rm F}$ have also been found for Sb$_2$Te$_3$, Bi$_2$Se$_3$, and Bi$_2$Te$_3$ after careful optimization of growth conditions in UHV \cite{Wang2010,Hoefer2014,Dai2016}.

\begin{figure*}[t]%
\includegraphics*[width=\textwidth]{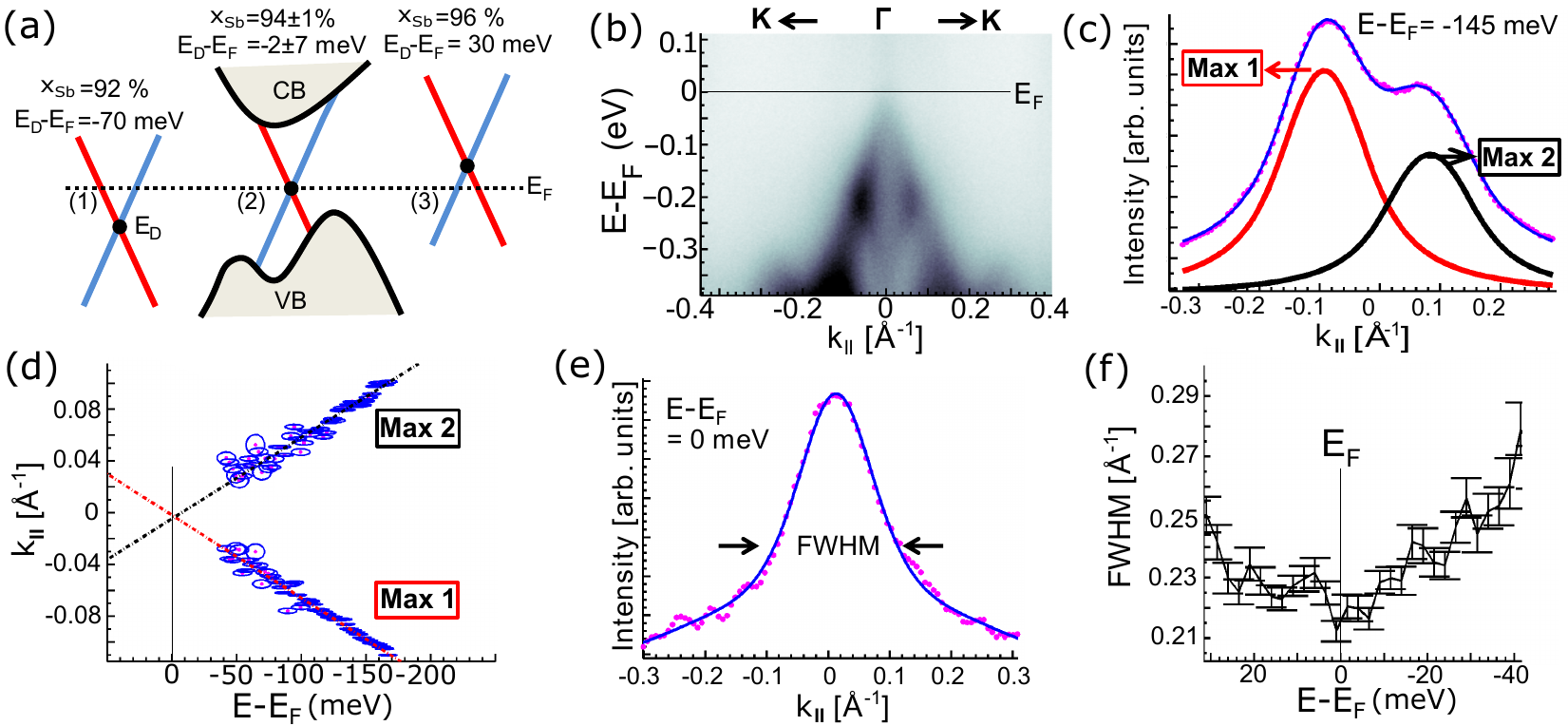}
\caption{{\it Tuning the Dirac point by stoichiometry.} (a) Model of Dirac cone (blue, red lines) between the valence band (VB) and conduction band (CB) marked as grey areas for  (Bi$_{1-x_{\rm Sb}}$Sb$_{x_{\rm Sb}}$)$_{2}$Te$_{3}$ at different $x_{\rm Sb}$. Resulting Dirac point energy $E_{\rm D}$ is indicated. (b) $E(k_{||})$ dispersion along $\overline{\mathrm K}\overline{\mathrm \Gamma}\overline{\mathrm K}$ (dark: high intensity, bright: low intensity), $h\nu=21.2$\,eV. Marked $E_{\rm F}$ is determined on polycrystalline Cu. (c) Cut through (b) (pink dots) at $E-E_{\mathrm{F}}=-145$\,meV with fit curve (blue) consisting of two Voigt curves (black, red lines with indicated peak positions Max 1, Max 2). (d) Energy dependent peak positions (Max 1, Max 2) deduced from fits as shown exemplarily in (c) (red points with surrounding ellipses that enclose the $2\sigma$ confidence area). A linear regression of the data points (black, red dashed lines) is used to determine  $E_{\mathrm{D}}-E_{\mathrm{F}}=-2\pm 7$\,meV. (e) Cut through (b) at $E=E_{\mathrm{F}}$ (pink dots) exhibiting only a single peak. One Voigt fit (blue) is used to deduce full width at half maximum (FWHM). (f) FWHM's of single peaks as in (e), i.e., close to $E_{\rm F}$. Error bars are marked. $T=50$\,K \cite{Kellner2015}.}
\label{Fig2}
\end{figure*}

Figure~\ref{Fig1}(e)$-$(g) show SARPES data recorded via a Mott detector, that probes the two in-plane directions of the spin. Two peaks at opposite ${\bm k}$ are recorded corresponding to the two opposite sides of the Dirac cone. The spin polarization is found to be exclusively perpendicular to the $\bm{k}$ vector as expected from the (Rashba-type) spin-orbit interaction. It is, moreover, helical, i.e., it switches sign when inverting $\bm{k}$. These are the typical fingerprints of a Dirac cone type TSS \cite{Zhang2009}. Out-of-plane spin polarizations have also been observed, in particular, further away from $E_{\rm D}$  and are traced back to distortions of the simple Dirac cone, e.g., via warping, i.e., by influences of the crystal structure \cite{Souma2011}. It is important to realize that SARPES does not probe the spin polarization of the initial state exclusively, but that the photoemission process is an excitation to unoccupied states extending into the vacuum that can change the spin polarization either by matrix effects or by spin polarization of the final state \cite{Huefner1995}. This can be captured by calculations within the so-called fully relativistic one-step model based on DFT calculations \cite{Ebert2011}. In particular, at low photon energies, it turns out that the detected spin polarization can even be inverted with respect to the initial state depending on the polarization direction of the exciting light \cite{Jozwiak2013}. At higher energies in the deep UV regime, this is less relevant, since excited states are well above the vacuum level.
Hence, the helicity of the TSS can be deduced  being counterclockwise for the lower part of the Dirac cone of Sb$_2$Te$_3$ (Fig.~\ref{Fig1}(d)). This is in accordance with the DFT calculations (Fig.~\ref{Fig1}(b)).
The absolute value of the spin polarization of the TSS is not extracted directly from our SARPES data due to the limited angular and energy resolution. The reduced resolution during SARPES with respect to ARPES is caused by the low efficiency of the Mott detector. Novel approaches improve this efficiency considerably via spin dependent, $\bm{k}$ conserving reflections of the photoelectrons at single crystals \cite{Medjanik2017}. Hence, resolution can be much better, but such apparatus was not available during the measurements presented in Fig.~\ref{Fig1}. Consequently, spin polarization had to be extracted rather indirectly by carefully subtracting the inelastic background, the background originating from the also measured spin-polarized surface states at lower energy (visible in Fig.~\ref{Fig1}(b) at $-0.4$ to $-0.8$\,eV), and the background from the overlapping, enclosed bulk states. Nevertheless, the accordingly best fit of the SARPES data revealed a spin polarization of the TSS of $80-95$\,\% (Fig.~\ref{Fig1}(g)) matching the DFT result of 90\,\% surprisingly well \cite{Pauly2012}.
Obviously, the TSS is not 100\,\% spin-polarized, albeit it is spin-helical. This is a natural consequence of spin-orbit interaction, that strongly mixes the spin with orbital degrees of freedom via the heavy atoms involved. Thus, spin is not a good quantum number in these materials.

\section{Tuning the Dirac point energy}
\begin{figure*}[tbh]%
\includegraphics*[width=\textwidth]{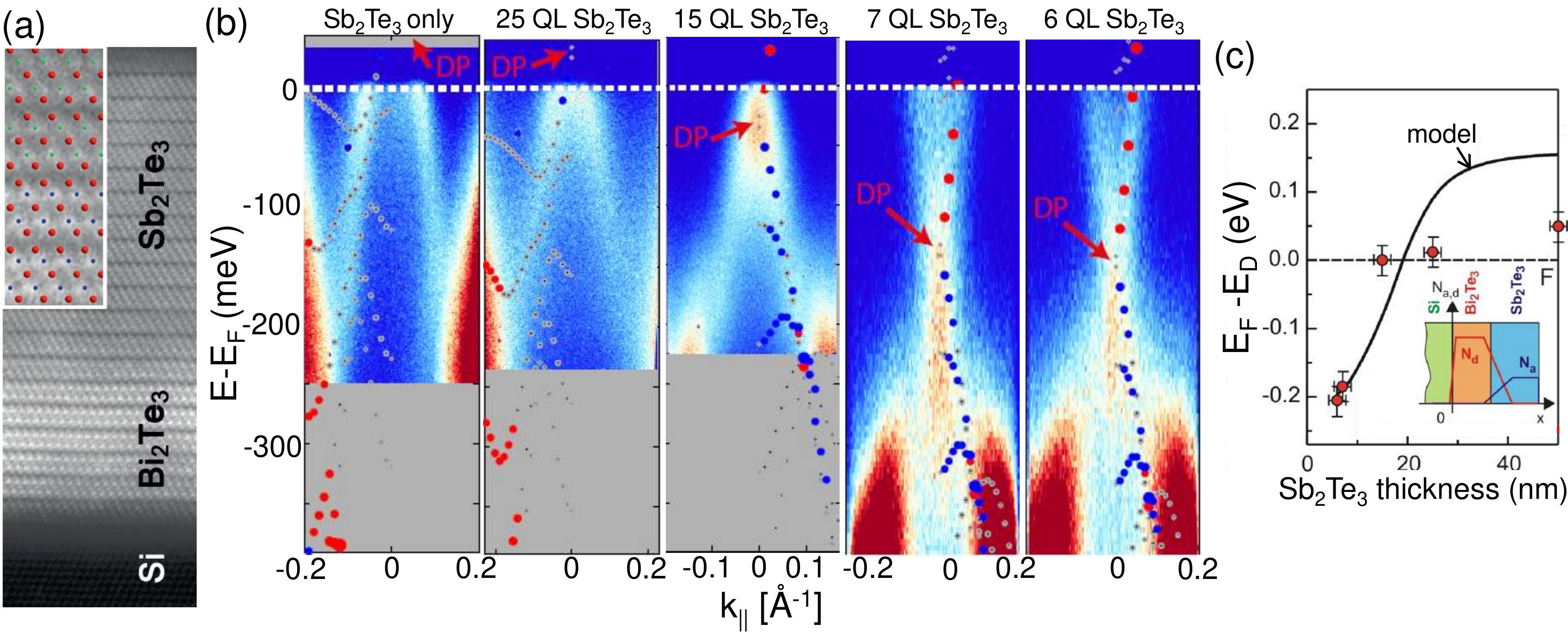}
\caption{{\it Tuning the Dirac point by band bending.} (a) High-angle annular dark-field (HAADF) cross section image recorded by TEM of Bi$_2$Te$_3$/Sb$_2$Te$_3$ stack grown by molecular beam epitaxy (MBE) on Si(111). Different brightness indicates chemical contrast via different atomic weights of Bi and Sb. Inset: Zoom with overlaid symbols representing Te atoms (red), Sb atoms (green) and Bi atoms (blue), $T=300$\,K. (b) ARPES data (blue-white-red color code) of different number of Sb$_2$Te$_3$ QLs (marked above) on top of Bi$_2$Te$_3$/Si, $h\nu=8.4$\,eV, $T=15$\,K. The first two (last three) data sets are recorded along $\overline{\rm K}\overline{\mathrm \Gamma}\overline{\rm K}$ ($\overline{\rm M}\overline{\mathrm \Gamma}\overline{\rm M}$). "Sb$_2$Te$_3$ only" marks thick Sb$_2$Te$_3$ on Si without Bi$_2$Te$_3$. DFT data of 6 QL Sb$_2$Te$_3$ (see also Fig.~\ref{Fig1}(b)) are overlaid (symbols) after shifting in energy in order to optimally match experimental data. The resulting Dirac point (DP) is marked. (c) Dirac point energies from ARPES (symbols) in comparison with a numerically solved one-dimensional Poisson-Schr\"odinger model  (full line). Inset: dopant densities used for the calculations as determined independently for both materials via Hall measurements (Bi$_2$Te$_3$: n-type $N_{\rm D}=2\cdot 10^{19}$\,cm$^{-3}$, Sb$_2$Te$_3$: p-type $N_{\rm A}=2\cdot 10^{18}$\,cm$^{-3}$). A linear interface intermixing across 5\,nm is included as deduced from Auger electron spectroscopy depth profiling.
 \cite{Eschbach2015}.}
\label{Fig3}
\end{figure*}
One main task after the experimental discovery of 3DTIs was to tune their Dirac point energy $E_{\rm D}$, that mostly turned out to be far away from $E_{\rm F}$ \cite{Ando2013}.
Hence, literally speaking, the first 3DTIs were not even insulators in their interior. More importantly, the transport properties of the 3DTIs could not be probed
without rendering the bulk of the material sufficiently insulating.
A rather obvious, initial approach was to exploit the opposite p-type doping of Sb$_2$Te$_3$ and n-type doping of  Bi$_2$Te$_3$ or  Bi$_2$Se$_3$.
Two main strategies have been pursued. Either, the two materials are mixed in a way such that they exhibit a similar density of acceptors and donors \cite{Ren2010,Kong2011}. This approach eventually led to the observation of the quantum Hall effect within thin films of BiSbTeSe$_2$ as a clear signature of dominating 2D-type transport \cite{Xu2014,Xu2016}. Detailed analysis of the filling factor dependence of the Hall conductance identified the TSSs on bottom and top surface as the origin of the half integer quantum Hall effect \cite{Xu2016}.
The respective tuning of the Dirac cone, respectively  $E_{\rm D}$, with respect to $E_{\rm F}$ can be monitored by ARPES in detail \cite{Zhang2011}. This is particularly important for the protection of Majorana states within vortices of a topological superconductor against conventional single-particle excitations by an effective gap $E_{\rm gap,eff}$ reading  \cite{Sau2010,Rakhmanov2011,Xu2014b}
\begin{equation}
E_{\rm gap,eff} \simeq \frac{\Delta^2}{\sqrt{\Delta^2+(E_{\rm F}-E_{\rm D})^2}}
\end{equation}
with $\Delta$ being the excitation gap of the surrounding topological superconductor.

\begin{figure*}[tbh]%
\includegraphics*[width=\textwidth]{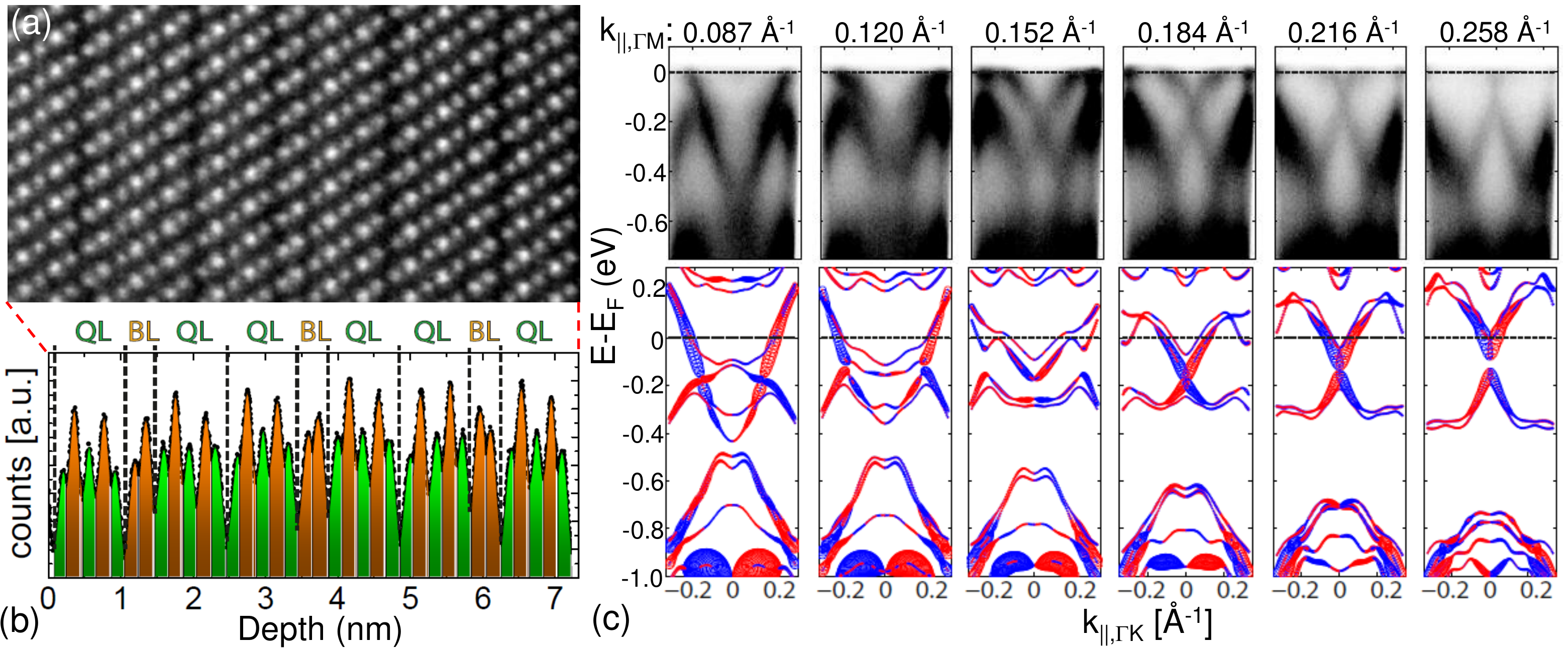}
\caption{{\it Dual topological insulator.} (a) Cross-sectional scanning TEM image of 39\,nm thick Bi$_1$Te$_1$ on Si(111) exhibiting atomic contrast caused by different atomic weights. (b) Intensity profile line of (a) along the horizontal direction after averaging along the vertical direction. Deduced Bi (Te) layers are colored yellow (green). Bi$_2$Te$_3$ quintuple layers (QL) and Bi bilayers (BL) are marked. (c) Upper row: ARPES data along  $\overline{\rm K}\overline{\mathrm \Gamma}\overline{\rm K}$ (dark: high intensity, bright: low intensity) for different $k_\parallel$ along $\overline{\mathrm \Gamma}\overline{\rm M}$ as marked on top, $h\nu=8.4$\,eV. Lower row: Corresponding band structure from DFT for a slab of 24 layers terminated by a single Bi$_2$Te$_3$ layer. Colors mark in-plane spin directions perpendicular to ${\bm k}_\parallel$ with spin polarization encoded as size of the circle. \cite{Eschbach2017}.}
\label{Fig4}
\end{figure*}

Figure~\ref{Fig2}(a) sketches the results for (Bi$_{1-x_{\rm Sb}}$Sb$_{x_{\rm Sb}})_2$Te$_3$ at different mixing of Sb and Bi including the case of $E_{\rm F} \simeq E_{\rm D}$. Corresponding ARPES data are displayed in
Fig.~\ref{Fig2}(b) \cite{Zhang2011,Kellner2015}. Figure~\ref{Fig2}(c)$-$(f) shows the evaluation of the data. The $k_\parallel$ values of the TSS are obtained from fitting intensity profiles $I(k_\parallel)$ (Fig.~\ref{Fig2}(c)) at different energies that are subsequently extrapolated linearly to determine the crossing point as $E_{\rm D}$ (Fig.~\ref{Fig2}(d)). Alternatively, the full width at half maximum (FWHM) of $I(k_\parallel)$ closer to $E_{\rm D}$ (Fig.~\ref{Fig2}(e)) is employed via identifying $E_{\rm D}$ as the energy with lowest FWHM (Fig.~\ref{Fig2}(f)). In both cases, $E_{\rm F}$ has to be carefully calibrated as well. For the particular sample, we found $E_{\rm F} \simeq E_{\rm D}$ within 5\,meV \cite{Kellner2015}. Since no time dependent band shifts were observed, the value is likely robust as long as the sample is in UHV. However, ex-situ Hall measurements on identically prepared samples exhibit a transition form p-type to n-type bulk conduction at much lower Sb concentration ($x_{\rm Sb}\simeq 60$\,\%) \cite{Weyrich2016}. Hence, rescuing the precise tuning for electric devices requires additional efforts and investigations.

Another approach uses the electric field at interfaces between p-type and n-type 3DTIs \cite{Wang2012,Eschbach2015}. As well known for semiconductor p-n junctions, a depletion
region forms at the interface such that a thin enough overlayer can maintain in the depletion region. This implies that $E_{\rm F}$ remains in the band gap up to the surface. The approach has the general advantage that it avoids ternary or quarternary alloys that potentially induce additional scattering centers for electrons via alloying.
Figure~\ref{Fig3}(a) displays a transmission electron microscope (TEM) image of a stack of n-type and p-type 3DTIs grown by molecular beam epitaxy (MBE). A relatively sharp interface is observed via the material contrast due to different atomic weights of Bi and Sb. Figure~\ref{Fig3}(b) shows ARPES data at different thickness of the upper p-type Sb$_2$Te$_3$ on n-type Bi$_2$Te$_3$. Obviously, the Dirac cone is shifted downwards in energy with decreasing Sb$_2$Te$_3$ thickness. This confirms the reasoning of an upwards $E_{\rm D}$ shift at the surface via the depletion zone. To determine $E_{\rm D}$ including the thicknesses, where it is above $E_{\rm F}$, DFT results of 6 QL Sb$_2$Te$_3$ are overlaid after rigidly shifting them to reproduce the ARPES data. It turned out that the best anchor point for shifting is the surface state at lower energy (Fig.~\ref{Fig1}(b) at $-0.4$ to $-0.8$\,eV). This state is vertically stronger confined to the surface area and, hence, is more intense in ARPES and less prone to the averaging by the vertical band bending \cite{Pauly2012} (details in \cite{Eschbach2015}). The resulting $E_{\rm D}-E_{\rm F}$ has been compared with the result of a 1D Poisson-Schr\"odinger model revealing reasonable agreement (Fig.~\ref{Fig3}(c)). The model is based on the charge carrier densities of  MBE grown films of Sb$_2$Te$_3$ and
Bi$_2$Te$_3$ as determined by Hall measurements, while assuming the same density of dopants and charge carriers. An intermixing at the interface is additionally taken into account that is deduced from Auger electron spectroscopy depth profiling \cite{Eschbach2015}.
Obviously, depletion method via p-n junction is also able to tune $E_{\rm F} \simeq E_{\rm D}$ for a thickness of $\sim  20$ QL Sb$_2$Te$_3$ on top of Bi$_2$Te$_3$.

\section{Materials with particular properties: Dual topological insulators and phase change materials}
\begin{figure*}[tbh]%
\includegraphics*[width=\textwidth]{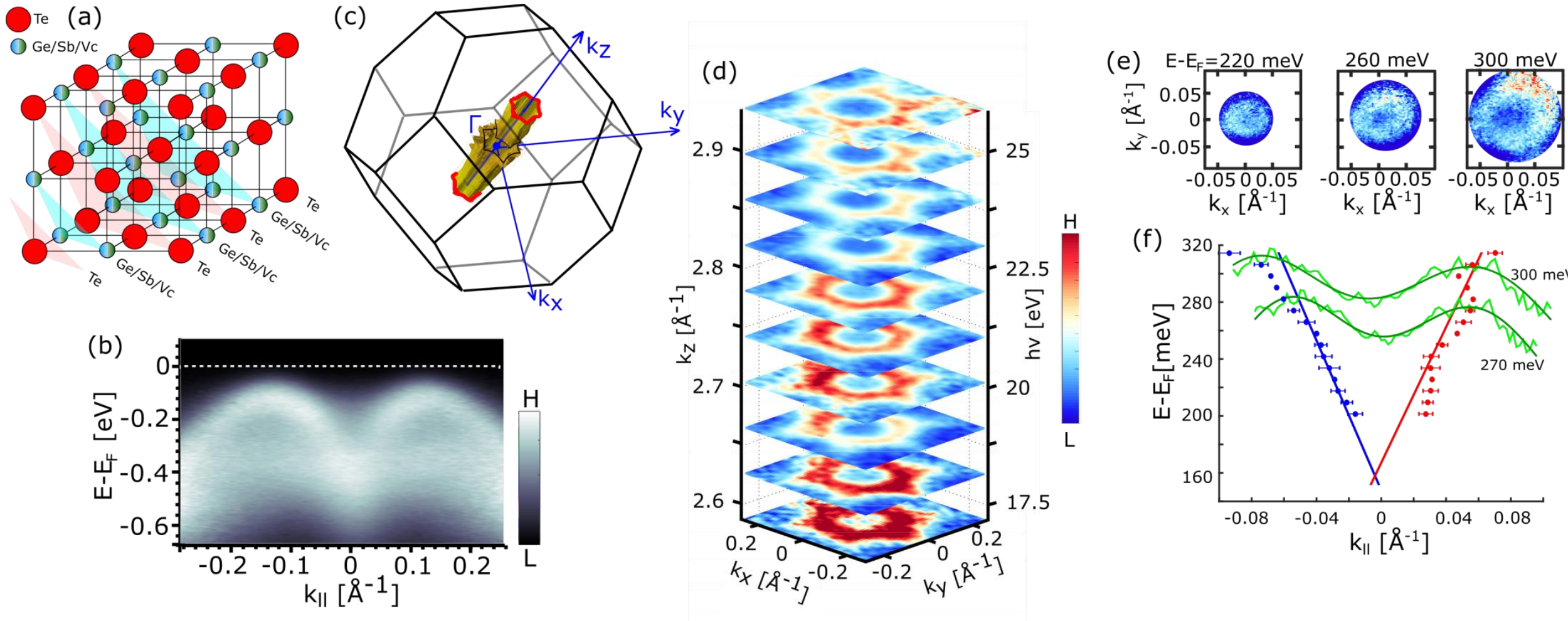}
\caption{{\it Topological phase change materials.} (a) Structural model of the metastable rocksalt structure of Ge$_2$Sb$_2$Te$_5$ with intermixed layers of Ge, Sb and vacancies (Ge/Sb/Vc). Adjacent (111) layers are highlighted by alternatingly colored, transparent triangles (pink, cyan). (b) ARPES data along $\overline{\rm K}\overline{\mathrm \Gamma}\overline{\rm K}$, $h\nu=22.5$\,eV. (c) Brillouin zone  of rock-salt Ge$_2$Sb$_2$Te$_5$  with principal $\bm {k}$ directions marked including the measured so-called pseudo Fermi surface (text) in gold. Resulting Fermi lines at the (111) side planes are drawn in red. (d) ARPES intensity across ($k_x$, $k_y$) plane at $E_{\rm F}$ recorded for different $h\nu$ as marked. The corresponding $k_z$ is calculated using an inner potential $E_{\rm inner} =14$\,eV as deduced from the symmetry of the ARPES data along $k_z$ (details in \cite{Kellner2018}). (e)  Two-photon ARPES intensity
$I(k_x, k_y)$ for different $E-E_{\rm F}$ above $E_{\rm F}$, i.e., within the unoccupied area of the band gap, pump $h\nu=1.63$\,eV, probe $h\nu=4.89$\,eV, time delay $\Delta t =1.33$\,ps. (f) Green lines: cuts through (e) along $k_y$  at $k_x=0$\,\AA$^{-1}$ (jagged lines) with fits consisting of two Voigt peaks (smooth lines). Red, blue dots: peak positions of corresponding Voigt fits for several energies after averaging the cuts along two perpendicular $\bm k$ directions. Red, blue lines: linear fits to the red and blue dots indicating a Dirac point at $E_{\rm D}-E_{\rm F}=160$\,meV. \cite{Kellner2018}.}
\label{Fig5}
\end{figure*}
\paragraph{Dual topological insulators} Another important application of ARPES is to confirm desired properties of novel topological materials. This includes topologically crystalline insulators (TCIs) \cite{Tanaka2012,Dziawa2012}, Dirac semimetals \cite{Liu2014,Liu2014b,Yan2017} and Weyl semimetals \cite{Xu2015,LV2015}. Interestingly, topological properties of different kind can be combined in a single material, if the topological indices belong to
different symmetries of the Hamiltonian \cite{Rauch2014}. For example, 3DTIs protected by time-reversal symmetry can be combined with TCIs protected by a crystal symmetry such as $n$-fold rotation or mirroring \cite{Fu2011}. This raises the perspective to break one of the symmetries, hence, switching between different topology types \cite{Rauch2014,Eschbach2017}.
The first material that experimentally showed dual topology was Bi$_1$Te$_1$ \cite{Eschbach2017}. It consists of stacked Bi bilayers (BLs) and Bi$_2$Te$_3$ QLs in a ratio of $1:2$ as evidenced by TEM (Fig.~\ref{Fig4}(a)$-$(b)). Bi BLs are well known to be 2DTIs \cite{Murakami2006,Yang2012,Drozdov2014} such that the stacking of such bilayers at sufficiently low interlayer interaction would result in a weak 3DTI. The so-called dark surface without TSS is simply the Bi BL surface, while the edge states of the BL lead to the TSSs at all other surfaces. The Bi$_2$Te$_3$ layers can be thought of as spacer layers between the Bi bilayers or as 2DTI layers themselves.
Indeed, DFT calculations find a small band gap of 0.1\,eV around $E_{\rm F}$ for the intrinsic, i.e., undoped, Bi$_1$Te$_1$ with topological indices (0;001). This indicates a weak 3DTI with its dark surface perpendicular to the (001) direction \cite{Eschbach2017}. However, the reasoning via stacked 2DTIs is too simple, when analyzing the DFT data in more detail. Interlayer hybridizations mix up the 2D bands strongly, such that the weak 3DTI properties are rather accidental and not directly related to the 2DTI properties of the constituting layers. Intriguingly, the mirror Chern number of the same gap around $E_{\rm F}$, that is protected by a mirror symmetry across the $(1\overline{1}00)$ plane, is $n_{\rm M}=-2$ rendering the system a TCI as well. Consequently, one expects an additional pair of Dirac cones on the dark (001) surface of the weak 3DTI Bi$_1$Te$_1$. The Dirac points of these Dirac cones are necessarily located on the line in ${\bm k}$ space where the (001) surface intersects with the $(1\overline{1}00)$ mirror plane. It must, moreover, be offset in opposite directions from $\overline{\mathrm \Gamma}$. Figure~\ref{Fig4}(c) (upper row) shows a set of ARPES data in $E(k_{\parallel,1})$ representation that are recorded perpendicular to this mirror line for increasing $k_{\parallel,2}$ values along the line. The data exhibit an apparent Dirac cone as crossing of two bands at $k_{\parallel,2} \simeq 0.18$\AA$^{-1}$ and $E-E_{\rm F} \simeq -0.2$\,eV.
The development of the bands with $k_{\parallel,2}$ towards the crossing agrees nicely with corresponding DFT results  of Bi$_1$Te$_1$(001) (Fig.~\ref{Fig4}(c), lower row). In order to achieve this agreement, the Bi$_1$Te$_1$ film had to be terminated by a single QL and had to be downshifted by 100\,meV with respect to $E_{\rm F}$, Both is reasonable with the latter accounting for n-type doping as expected from the well known n-type doping of Bi$_2$Te$_3$. The good agrement between ARPES and DFT data, also found for multiple other bands of Bi$_1$Te$_1$, is the central evidence for the dual topological character of Bi$_1$Te$_1$ \cite{Eschbach2017}.

\begin{figure*}[tbh]%
\includegraphics*[width=\textwidth]{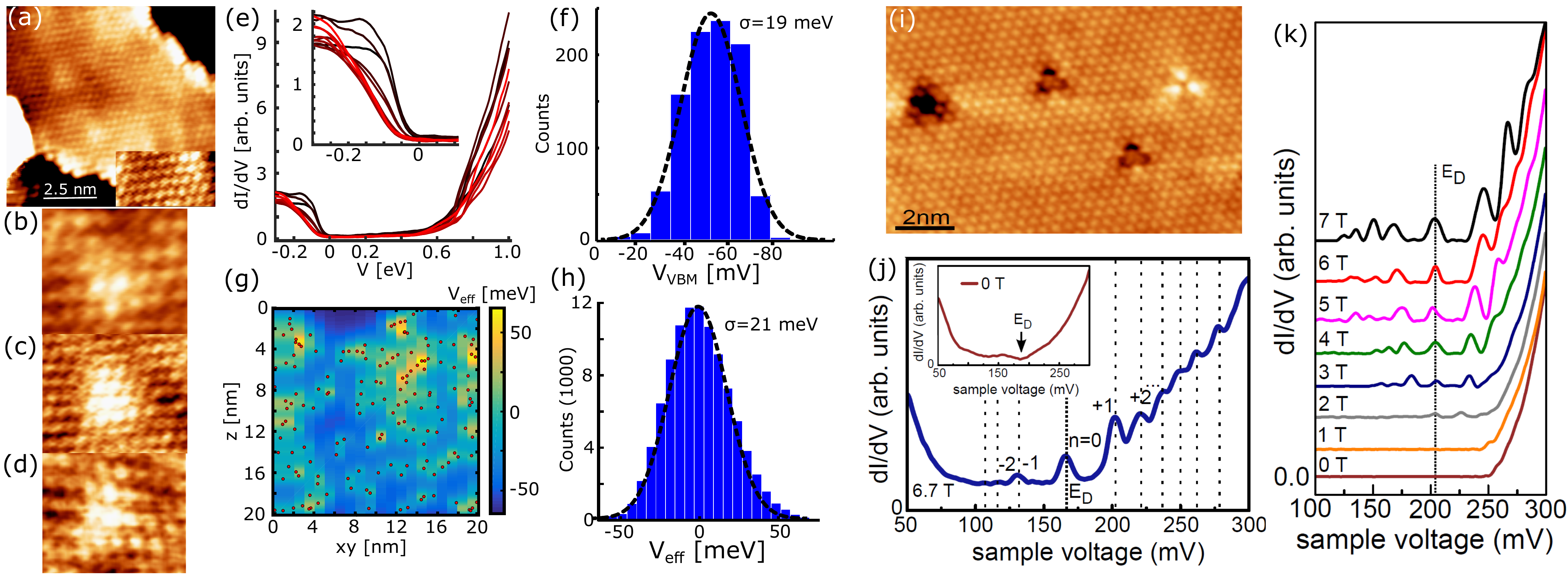}
\caption{{\it Mapping the disorder potential.} (a) STM image of in-situ transferred Ge$_2$Sb$_2$Te$_5$(111) grown by MBE on Si(111). The hexagonal atomic structure of the top Te layer is visible with inset at larger magnification, $V=-0.5$\,V, $I=100$\,pA, $T=300$\,K. (b)$-$(d) STM images of characteristic triangular protrusions indicating subsurface defects, $V=-0.5$\,V, $I=100$\,pA, $T=300$\,K.
(e) Scaled $dI/dV(V)$ recorded at adjacent locations, $V_{\rm stab}=-300$\,mV, $I_{\rm stab}=50$\,pA, $T=9$\,K. Inset: zoom into the region of the valence band maximum (VBM). (f) Histogram of valence band onsets $V_\mathrm{VBM}$ as deduced from the peak energies in $d^3I/dV^3 (V)$ curves (blue bars). A dashed Gaussian fit with marked $\sigma$-width is added. (e) Vertical cut through the simulated electrostatic potential $V_{\mathrm{eff}}(x,y,z)$ for randomly distributed bulk acceptors (red dots) at density $N_{\rm A}=3\cdot 10^{26}$/m$^3$ as deduced from Hall measurements. (h) Histogram of the potential values $V_{\mathrm{eff}} (x,y)$ at the surface resulting from multiple simulations as in (g) (blue bars). A dashed Gaussian fit with marked $\sigma$-width is added. (i) STM image of in-situ cleaved Sb$_2$Te$_3$(0001) exhibiting the hexagonally arranged top Te layer with clover-shaped defects likely Sb$_{\rm Te}$ (bright) and subsurface Vac$_{\rm Sb}$ (dark), $V=0.4$\,V, $I=1$\,nA, $T=6$\,K. (j) $dI/dV(V)$ at $B=6.7$\,T showing Landau levels of the topological surface state marked by level index $n$ at the dashed lines that result from Lorentzian fits of the peaks, $V_{\rm stab}=0.3$\,V, $I_{\rm stab}=400$\,pA, $V_{\rm mod}=4$\,mV, $T=6$\,K. $E_{\rm D}$ is located at Landau level $n=0$. Inset: $dI/dV(V)$ at $B=0$\,T (same position) with $E_{\rm D}$ marking the minimum of the curve, $V_{\rm stab}=0.3$\,V, $I_{\rm stab}=50$\,pA, $V_{\rm mod}=4$\,mV, $T=6$\,K. (k) $dI/dV(V)$ at $B = 0 -7$\,T as marked recorded on a different sample area as (j) and offset vertically, $V_{\mathrm{stab}}=0.3$\,V, $I_{\mathrm{stab}}=100$\,pA, $V_{\mathrm{mod}}=2$\,mV, $T=6\,K$. The vertical dotted line indicates Landau level $n=0$, hence, $E_{\rm D}$. (a)$-$(h) Ge$_2$Sb$_2$Te$_5$ \cite{Kellner2017}, (i)$-$(k) Sb$_2$Te$_3$ \cite{Pauly2015}.}
\label{Fig6}
\end{figure*}

\paragraph{Topological phase change materials} Another interesting class of 3DTI materials are commercially used in electronic applications. They are called phase change materials (PCMs) providing two favorable properties for data storage. Firstly, they are fastly switchable (ns-scale) between the amorphous and a metastable crystalline phase at low energy penalty \cite{Xiong2011,Loke2012}. Secondly, they exhibit a strong difference in optical reflectivity and electric conductivity between the two phases \cite{Wuttig2012}.
Consequently, they are employed in DVDs and random-access memories \cite{Wuttig2012}. A standard class of PCMs is found on the pseudobinary line between the strong 3DTI Sb$_2$Te$_3$ and GeTe \cite{Wuttig2007}. The later material is also strongly influenced by spin-orbit interaction revealing a strongly Rashba-split
surface state and a strongly Rashba-split bulk state at $E_{\rm F}$ \cite{Liebmann2015,Elmers2016,Krempask2016}. Hence, it is natural to assume that some of the PCMs are 3DTIs as well.
Indeed, DFT predicts 3DTI properties \cite{Kim2010,Eremeev2012,Kim2012,Silkin2013} and finds that the 3DTI character depends strongly on details of the atomic arrangement \cite{Kim2012,Silkin2013}. Figure~\ref{Fig5}(a) shows the structure of the most commonly used PCM Ge$_2$Sb$_2$Te$_5$ in its metastable phase. It consists of alternating layers of hexagonal Te and a mixture of Ge, Sb and vacancies. If these ABC stacked layers exhibit additional order in the Ge/Sb/vacancy layers depends on details of the preparation and is decisive for the 3DTI properties according to DFT \cite{Kim2010,Eremeev2012,Kim2012,Silkin2013}. Hence, a subtle borderline between
strong 3DTI and trivial properties appears. Based on these findings, it has even been speculated that the reversible, strong difference in electric conductivity of a superlattice GeTe/Sb$_2$Te$_3$, that appears after applying voltage pulses without making the material amorphous, is caused by a switch in topology \cite{Simpson2011,Tominaga2013}.

However, the experimental confirmation of strong 3DTI properties in PCMs is complicated by the p-type doping. Hence, initially the only evidence by ARPES was the M shaped valence band with maxima away from $\overline{\mathrm \Gamma}$ (Fig.~\ref{Fig5}(b)) \cite{Pauly2013}. This configuration has been found in DFT calculations only for inverted bands representing a strong 3DTI \cite{Kim2010,Eremeev2012,Kim2012,Silkin2013}. The p-type character of these bands is unconventional, since the energetically highest peak position in $I(E)$ plots is about 100\,meV below $E_{\rm F}$ (Fig.~\ref{Fig5}(b)). The p-type doping is instead realized by the tails of the disorder broadened valence bands that cut $E_{\rm F}$.
 Hence, the ARPES intensity at $E_{\rm F}$ provides a $\bm k$ distribution mimicking the highest energy peaks below $E_{\rm F}$ (Fig.~\ref{Fig5}(c), (d)). This is called the pseudo Fermi surface that can indeed be used to deduce the charge carrier density in good agreement with Hall measurements, if adequately weighted by the broadening of the valence bands due to disorder \cite{Kellner2018}.
  Hence, disorder in the Ge/Sb/vacancy layer is the central ingredient for the conductivity of the metastable PCM phase \cite{Kellner2018}.

  The TSS above the broadened valence bands has been found by two-photon ARPES, i.e., a first light pulse transfers electrons into the initially unoccupied TSS and a second light pulse with time delay $\Delta t$ extracts photoelectrons from the now occupied TSS. Figure~\ref{Fig5}(e) shows data for several energies above $E_{\rm F}$ exhibiting a rather isotropic circle in $\bm k$ space. The circle shrinks in diameter with decreasing energy. Extrapolation of the radius to lower energies (Fig.~\ref{Fig5}(f)) implies vanishing diameter at about 160\,meV above $E_{\rm F}$ that represents $E_{\rm D}$. Hence, the well-established conducting phase of the PCM  Ge$_2$Sb$_2$Te$_5$ is a strong 3DTI, at least, after the preparation by MBE as probed in this study \cite{Kellner2018}. This is appealing for 3DTI-based applications via exploiting the established expertise for upscaling conventional Ge$_2$Sb$_2$Te$_5$ devices \cite{Hayat2017}. Counteracting the unfavorable p-doping of Ge$_2$Sb$_2$Te$_5$ is possible by replacement of Ge with the heavier Sn \cite{Schaefer2017}, where, however, 3DTI properties still have to be demonstrated experimentally \cite{Eremeev2015}.

\section{Disorder characterization}

As described in the introduction, a central task for improving the electric transport properties of 3DTIs (and 2DTIs) is the reduction of disorder. Disorder can lead to additional transport channels concealing the features of the TSS as well as to scattering of the TSS electrons \cite{Ando2013,Culcer2012}. STS is the tool of choice for probing the disorder at the surface due to its unprecedented spatial and energy resolution in probing the LDOS. It has only the minor drawback that it is exclusively measuring the surface disorder and not the disorder within deeper layers of the bulk of the crystal \cite{Zandvliet2009}.

One possibility by STS is to track characteristic features of the energy dependent LDOS \cite{Beidenkopf2011,Dai2016}. One measures $dI/dV(V)$ curves with $I$ being the tunnel current and $V$ being the voltage applied between tip and sample. Mostly, such curves are measured by lock-in technique, i.e., the tip-surface distance is stabilized at voltage $V_{\rm stab}$ and current $I_{\rm stab}$. Afterwards, the feedback loop is switched off, such that the tip surface distance remains constant, while the voltage is changed linearly and overlapped with an oscillating voltage of amplitude $V_{\rm mod}$ that enables the phase sensitive detection of $dI/dV$ via a lock-in amplifier.  In first order, the resulting $dI/dV(V)$ represents the LDOS$(E-E_{\rm F})$ \cite{Tersoff1985,Morgenstern2000b,Morgenstern2003,Zandvliet2009}. This gives direct access, e.g., to spatial variations of the band gap for a semiconductor or insulator.

Figure~\ref{Fig6}(a) shows the (111) surface of the strong 3DTI Ge$_2$Sb$_2$Te$_5$ exhibiting Te as the top layer with hexagonal atomic structure \cite{Kellner2017}. Several, largely triangular bright protrusions appear on top of the atomic lattice (Fig.~\ref{Fig6}(b)$-$(d)). They have been identified as subsurface defects by comparison with DFT data \cite{Kellner2017}. The lateral size of the triangle increases with the depth of the defect below the surface. The particular sample grown by MBE exhibits a defect density of $\sim 1.5\cdot 10^{12}$/cm$^2$. This implies a potential disorder due to the positive charging of most of the defects, in particular, vacancies \cite{Pauly2013,Kellner2017,Wuttig2006}. The $dI/dV(V)$ curves (Fig.~\ref{Fig6}(e)) show a band gap of about 0.5\,eV with the valence band onset being close to $E_{\rm F}$ in agreement with optical absorption \cite{Lee2005} and ARPES data (Fig.~\ref{Fig5}(b)), respectively. The band gap onset is spatially varying. It is quantified via the peak position of the numerically determined $dI^3/dV^3(V)$ curves leading to a nearly Gaussian distribution of the spatially varying valence band onset with $\sigma$ width of $~20$\,meV (Fig.~\ref{Fig6}(f)). We compare this with a simple model calculation randomly distributing positive point charges with a density identical to the charge carrier density determined by Hall measurements (Fig.~\ref{Fig6}(g)). This leads to potential fluctuations on the surface with the same $\sigma$ width as in the experiment (Fig.~\ref{Fig6}(h)). It implies that the Coulomb centers of the charged acceptors (vacancies) dominate disorder in this sample. Interestingly, the LDOS does not vanish within the band gap (Fig.~\ref{Fig6}(e)) indicating the presence of in-gap surface states in agreement with the two-photon ARPES revealing a TSS (Fig.~\ref{Fig5}(f)).

Another possibility to map potential disorder is Landau level spectroscopy, however, requiring a magnetic field. It exploits the Dirac type spin chirality of the TSS implying a so called zeroth Landau level (LL0) that is tied to $E_{\rm D}$ \cite{Novoselov2005,CastroNeto2009}. Hence, tracking LL0 across the surface maps the potential disorder as seen by the TSS, i.e., averaged across some of the upper QLs \cite{Okada2012,Fu2014,Pauly2015}. The lateral spatial resolution of the method is largely given by the magnetic length \cite{Champel2010}. Figure~\ref{Fig6}(i) shows STM data of in-situ cleaved
Sb$_2$Te$_3$(0001) featuring a few defects that have been identified previously by comparison with DFT calculations as Sb substitutional in the upper Te layer (Sb$_{\rm Te}$, bright)
and vacancies in the Sb layer directly below the surface (Vac$_{\rm Sb}$, dark) \cite{Jiang2012}. We find a defect density of $4\cdot 10^{12}$/cm$^2$ with all apparent defects attributed to the upper QL \cite{Pauly2015}. Figure~\ref{Fig6}(j)-(k) show Landau level spectra recorded at two different locations of the sample. It is apparent that the energy of LL0 does not shift with $B$ field (Fig.~\ref{Fig6}(k)). Moreover, LL0 appears at the same energy as the minimum in $dI/dV(V)$ curves at $B=0$\,T. Finally, LL0 deviates by $\sim 40$\,meV between the two probed areas indicating the potential fluctuations.
We found that the deduced LL0 energy correlates with the local density of defects visible in the STM data (not shown) \cite{Pauly2015}.

\section{Edge states of weak topological insulators}

\begin{figure*}[tbh]%
\includegraphics*[width=\textwidth]{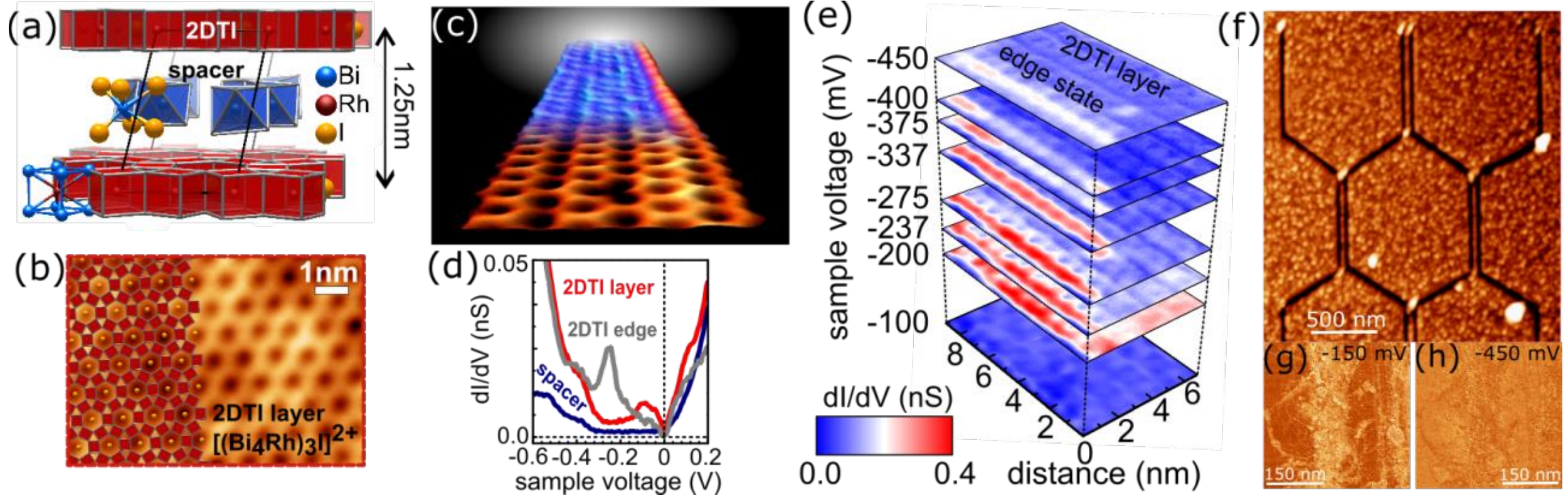}
\caption{{\it Edge states on the dark side of a weak topological insulator.} (a) Atomic model of Bi$_{14}$Rh$_3$I$_9$ consisting of alternating layers of the
2D topological insulator (Bi$_4$Rh)$_3$I (red) and the trivial insulator Bi$_2$I$_8$ (blue). (b) STM image recorded on a (Bi$_4$Rh)$_3$I terrace with atomic model structure overlaid using the same color code as in (b), $V = 1.5$\,V, $I = 100$\,pA. (c) 3D perspective of adjacent STM image (front area, $V = 0.8$\,V, $I = 100$\,pA) and $dI/dV$ image (background, $V=-0.337$\,V, $I=100$\,pA, $V_{\rm mod}=4$\,mV), both recorded at the same (Bi$_4$Rh)$_3$I terrace confined by a step edge on the right. (d)  $dI/dV(V)$ recorded at a step edge of the (Bi$_4$Rh)$_3$I layer (grey), on a (Bi$_4$Rh)$_3$I terrace (red) and on a Bi$_2$I$_8$ terrace (blue), $V_{\rm stab}=0.8$\,V, $I_{\rm stab}=100$\,pA, $V_{\rm mod}=4$\,mV. Notice the linearly vanishing $dI/dV$ intensity around $E_{\rm F}$ caused by an Efros-Shklovskii type Coulomb gap \cite{Efros1975}. (e) Stacked $dI/dV$ images recorded at the step edge of a (Bi$_4$Rh)$_3$I terrace for different $V$ across the band gap as marked on the left, $I_{\rm stab}=100$\,pA, $V_{\rm mod}=4$\,mV. (f) Tapping-mode AFM image of Bi$_{14}$Rh$_3$I$_9$ after scratching a network of step edges into the surface by a carbon coated Si cantilever at force $F=1$\,${\mathrm \mu}$N, $f_{\rm res}=275$\,kHz, $k=43$\,N/m, $A=30$\,nm, set point: 70\,\% (details \cite{Pauly2015b}). (g), (h) $dI/dV$ images of  a scratch accomplished by one STM tip within UHV and afterwards recorded by another tip at $V$ as marked, $I=300$\,pA, $V_{\rm mod}=20$\,mV. (a)$-$(f) $T=6$\,K \cite{Pauly2015b}, (g), (h) $T=300$\,K.}
\label{Fig7}
\end{figure*}
Weak 3DTIs have initially barely been studied due to the wrong conjecture that they are unstable with respect to most type of perturbations \cite{Fu2007}.
More detailed studies, however, revealed that the only detrimental perturbation is a strong dimerization of adjacent layers along the surface normal of the dark surface leading to a doubling of the unit cell \cite{Ringel2012,Obuse2014}. Hence, also weak 3DTIs typically exhibit robust spin-helical surface states protected from backscattering. The most simple way to construct a weak 3DTI is stacking 2DTIs without interlayer interaction \cite{Fu2007,Obuse2014}. This naturally implies that single-layer terraces on the dark surface are patches of 2DTIs that consequently must host one-dimensional topological edge states at its step edges. These edge states are spin helical and, hence, ideal conductors as long as time-reversal symmetry is not broken \cite{Yoshimura2013}. It turns out that such edge states appear generally for weak 3DTIs even if constructed differently \cite{Yoshimura2013}. This implies the possibility to scratch a network of ideal conductors into the surface of a weak 3DTI \cite{Pauly2015b}.

The first experimental realization of a weak 3DTI was Bi$_{14}$Rh$_3$I$_9$ \cite{Rasche2013}. It consists of alternating layers of the 2DTI (Bi$_4$Rh)$_3$I \cite{Rasche2016} and the trivial insulator Bi$_2$I$_8$ (Fig.~\ref{Fig7}(a)). The 2DTI exhibits a honeycomb unit cell such as graphene, but is made of the heavy atoms Bi, I and Rh (Fig.~\ref{Fig7}(b)). It, thus, mimics the initial idea of a 2DTI in a honeycomb lattice \cite{Kane2005}, but provides a much stronger spin-orbit interaction ($\sim 1$\,eV) leading to a sizable inverted band gap of $200-300$\,meV \cite{Rasche2016}. This gap is much larger
than in graphene with inverted band gap of $\sim 20$\,$\mu$eV \cite{Konschuh2010}. Hence, the idea to construct the 3D material is to stack 2DTI honeycomb structures \cite{Kane2005} that are separated by trivial insulators as spacers impeding interactions between the 2DTI layers.
However, it turned out that the strong spin-orbit interaction shifts much more bands across $E_{\rm F}$ than only the initial Dirac cone of the honeycomb lattice that appears at $E_{\rm F}$ without spin-orbit interaction \cite{CastroNeto2009}. Thus, the topological indices of a weak 3DTI again appear rather accidentally via inversion of several bands at the TRIMs of the Brillouin zone \cite{Rasche2016}.
Nevertheless, topological edge states at each step edge are expected and have been found by STS. They are directly visible as enhanced LDOS  intensity at step edges (Fig.~\ref{Fig7}(c), background). In $dI/dV(V)$ curves, the band gap region of the material ($-0.15$ to $-0.35$ \,eV) exhibits strong intensity exclusively at the step edges (Fig.\ref{Fig7}(d)). The edge states appear continuously along all edges \cite{Pauly2015b} and are only $\sim 1$\,nm wide perpendicular to the edge (Fig.~\ref{Fig7}(e)). Moreover, the edge states did not exhibit any fingerprints of standing waves, but only intensity modulations periodic with the unit cell as expected for Bloch states. Thus, backscattering is largely impeded. Networks of topological edge states can indeed be scratched into the surface either by the tip of an atomic force microscope (AFM) (Fig.~\ref{Fig7}(f)) with separation down to 25 nm \cite{Pauly2015b} or by the tip of an STM. The resulting scratches indeed show an increased LDOS within the band gap (Fig.\ref{Fig7}(g)), but not at energies outside the gap (Fig.~\ref{Fig7}(h)).

\begin{figure*}[tbh]%
\centering \includegraphics*[width=0.8\textwidth]{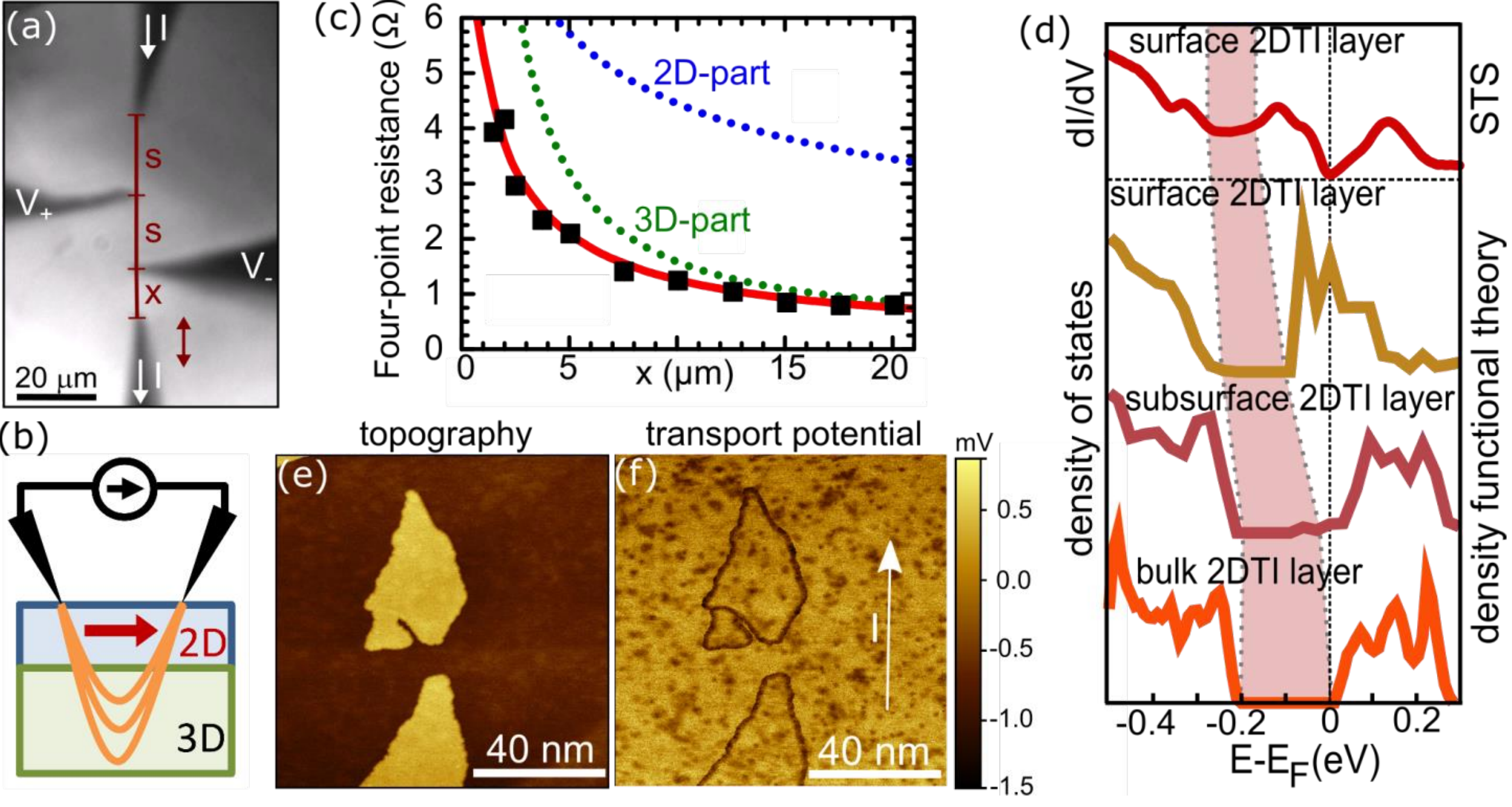}
\caption{{\it In-situ electric transport of weak topological insulator Bi$_{14}$Rh$_3$I$_9$.} (a) Optical microscope image of the four tips positioned on Bi$_{14}$Rh$_3$I$_9$ within UHV.
While the distances $s=20$\,${\rm \mu}$m are kept constant during the four-point resistance measurements, $x$ is changed. Applied current $I$ and measured voltages $V_+$ and $V_-$ are marked. (b) Sketch of the two parallel transport paths (2D, 3D) used for the simulations of the four-point resistance in (c). (c) Measured resistance $R=\frac{V_+-V_-}{I}$ as function of $x$ (black squares), $I=0-300$\,${\mathrm \mu}$A used to extract the resistance from the resulting linear $V(I)$ curve. A fit curve (red) is added assuming both, the  conductance of a 3D bulk contribution (green) and a 2D-type (blue) contribution \cite{Just2015,Luepke2018}. (d) Top curve: $dI/dV(V)$ recorded on a 2DTI layer ((Bi$_4$Rh)$_3$I), $V_{\rm stab} = 1$\,V, $I_{\rm stab} = 70$\,pA, $V_{\rm mod} = 4$\,mV. Two middle curves: layer resolved density of states of the surface and the inner 2DTI layer as deduced from a DFT calculation (FPLO code \cite{Koepernik1999}) of a slab with alternatingly stacked two 2DTI layers and two spacer layers (Bi$_2$I$_8$). Bottom curve: layer resolved density of states of the 2DTI layer deduced from a DFT calculation of an infinitely extended bulk crystal. Gap areas are marked in pink for all 4 curves \cite{Pauly2016}. (e) STM image of in-situ cleaved Bi$_{14}$Rh$_3$I$_9$, $V=-0.1$\,V, $I=240$\,pA. (f) Scanning tunneling potentiometry image of the same area as (e) with indicated direction of applied current, $I=1.2$\,mA, distance of current carrying tips along vertical direction: 7.5\,${\mathrm \mu}$m.}
\label{Fig8}
\end{figure*}
Unfortunately, $E_{\rm F}$ is not within the band gap and, thus, the edge states are not accessible by electric transport. Four-tip STM measurements in UHV (Fig.~\ref{Fig8}(a)) \cite{Voigtlaender2018}, however, revealed that the resistance as a function of distance between the tips is not described by a 3D transport model only, but required a sizable contribution from a parallel 2D transport channel (Fig.~\ref{Fig8}(b)$-$(c)).
The best fit of the experimental data (red curve in Fig.~\ref{Fig8}(c), \cite{Just2015,Just2017}) implies conductances for the 2D and 3D contribution $\sigma_{\rm 2D}=0.064\pm 0.005$\,S and $\sigma_{\rm 3D}=9200\pm 800$\,S/m, respectively. Thus, the 2D conductance corresponds to a $\sim 7$\,${\rm \mu}$m thick layer with the 3D conductance $\sigma_{\rm 3D}$.

This implies that the surface region of Bi$_{14}$Rh$_3$I$_9$ is significantly more conductive than the
bulk. The encouraging finding is corroborated by DFT calculations of bulk Bi$_{14}$Rh$_3$I$_9$ (Fig.\ref{Fig8}(d), bottom, orange curve) showing $E_{\rm F}$ within the band gap. Additional calculations of a thin film revealed that the surface is strongly n-doped (Fig.\ref{Fig8}(d), yellow curve) with the band gap at similar energies as found in the STS data (Fig.\ref{Fig8}(d), top, red curve) \cite{Pauly2016}. This  is in line with the strong 2D conductivity found by 4-tip STM. The band gap favorably moves quickly towards its bulk position already for the subsurface layer (Fig.\ref{Fig8}(d), pink curve). To explain the surface n-doping, we consider the charging of the individual layers. It turns out that the 2DTI layer (Bi$_4$Rh)$_3$I transfers about one electron per unit cell to each of its neighboring spacer layers Bi$_2$I$_8$ such that it is positively charged by about 2 electrons per unit cell in equilibrium. Under these circumstances, $E_{\rm F}$ is in the band gap of the 2DTI layer.
At the surface, however, one neighboring spacer layer is missing, such that about one electron per unit cell remains on the 2DTI layer making it strongly n-doped  \cite{Pauly2016}. In principle, this could be counteracted by adding acceptors such as iodine onto the surface, but a relatively large amount of about one iodine atom per unit cell is required \cite{Ghimire2017}.

Using the four-tip STM, we also performed scanning tunneling potentiometry \cite{Muralt1986}. This method measures the tip voltage $V$ that is required to nullify the current between tip and sample. Consequently, it maps the local potential, typically while current is flowing laterally. With four-tip STM, two tips can be used to inject the current, while a third tip is scanned in between to probe the nullifying voltage \cite{Luepke2015,Voigtlaender2018}. Consequently, the current induced potential is mapped. Figure~\ref{Fig8}(e) shows an area of the surface with 2DTI only, i.e., the islands exhibit step edges with height of a
combined 2DTI and spacer layer. The potentiometry data show a barely visible overall decrease of the potential from the bottom to the top by about 0.1\,mV due to the transport resistance. Much stronger features appear at the step edges and as patches on the surface of the 2DTI layers. They are identically present without applying current and are, hence, a static feature of the
surface. Such features are caused by thermo-voltage $V_{\rm thermo}$
resulting from a temperature difference of tip and sample $\Delta T \simeq 1$\,K and, as such, indicate spatially different slopes of the LDOS at
$E_{\rm F}$  according to $V_{\rm thermo} \propto \frac{d\ln{(LDOS(E))}}{dE}|_{E_{\rm F}}$ \cite{Stovneng1990,Druga2010}. As a result, the method reveals fluctuations of LDOS($E_{\rm F}$) on the 2DTI terraces and a significant difference between step edges and terraces. The experiment did not provide any indication of preferred transport along the step edges in agreement with the observation that the topological edge states are not at $E_{\rm F}$.

Other weak 3DTIs have been found \cite{Li2016,Wu2016,Hosen2018,Noguchi2019}, but none with $E_{\rm F}$ in the topological band gap. Some uncertainty remains for ZrTe$_5$, that is very close to a topological phase transition such that details on strain and temperature change the topological properties partly also in a favorable way \cite{Zhang2017}. More interestingly, bismuthene, a honeycomb Bi monolayer on SiC(0001), is a 2DTI that can be prepared in UHV with $E_{\rm F}$ inside the topological band gap of size $\sim 1$\,eV \cite{Reis2017}. Here, preferential transport along step edges might be detected by four-tip scanning tunneling potentiometry. Also the ideal conductance of the edge state could be probed. It would lead to a potential drop that only appears at the end of the step edge, i.e., at the transition to the terrace in current direction \cite{Klass1992}.

\section{Conclusions}

In this article, we summarized some of the key contributions of surface science methods to the development of 3DTIs. Most importantly, ARPES could identify strong 3DTIs via the Dirac cone and its spin helicity of the topological surface state, while STS could identify weak 3DTIs via their helical edge states protected from backscattering  at the dark surface.
Moreover, ARPES was instrumental to monitor the tuning of the Dirac cone towards $E_{\rm F}$, albeit the results are not compatible yet with the results from electric transport likely due to different treatment of the surfaces. Complementary, STS can map the potential disorder, most precisely via Landau level spectroscopy, and, hence, can monitor efforts to improve sample homogeneity. We have also shown exemplarily that particularly interesting materials can be identified as topological. In detail, we have discovered the first dual 3DTI Bi$_1$Te$_1$ and strong 3DTI properties in phase change materials as an example material used in commercial applications. Two-photon ARPES was crucial to find the Dirac cone in these materials that only appeared in the occupied states due to strong p-doping.
Finally, we have introduced the abilities of four-tip STM that can provide electric transport data in UHV without the requirement of ex-situ contacting. We anticipate that this method will be perspectively important to adapt the results from ARPES and STS to electric transport and, hence, to devices, since different surface treatments, that lead, e.g., to contaminations due to lithography, can be avoided.

\section{acknowledgement}
We strongly appreciate the previous contributions to the publications reviewed in this manuscript by P. Bhaskar, S. Bl\"ugel, S. Borisenko, J.E. Boschker,  V. Bragaglia, R. Callarco, S. Checchi, N. Demarina, V. L. Deringer, S. D\"oring,  R. Dronskowski, Th. Fauster, M. Gehlmann, A. Georgi, E. Golias, P. Gospodaric, M. Grob, D. Gr\"utzmacher,  A. Guissani, C. Holl, B. Holl\"ander, J. Kampmeier, B. Kaufmann, K. Koepernik, P. K\"uppers, M. Lanius, M. Luysberg, E. Mlynczak, E. Neumann, C. Niu, O. Rader,  J. Sanchez-Barriga, T. Sch\"apers, C.M. Schneider, M.R. Scholz, P. Sch\"uffelgen, D. Subramaniam, J. van den Brink, A. Varykhalov, R. N. Wang, and C. Weyrich.
We, moreover, gratefully acknowledge funding by the Deutsche Forschungsgemeinschaft (DFG, German Research Foundation) via the project Mo 858/13-2 within the priority programme SPP1666 "Topological Insulators" and via the Strategy Clusters of Excellence "Matter and Light for Quantum Computing (ML4Q)" EXC 2004/1 – 390534769 as well as "Complexity and Topology in Quantum Matter (ct.qmat)" ExC 2147, project-id 39085-490.




\end{document}